\documentclass[sigconf]{acmart}

\usepackage{subcaption}
\usepackage{url}
\usepackage[neverdecrease]{paralist}
\usepackage{graphics}
\usepackage{tabularx}
\usepackage{xspace} 
\usepackage{color,soul}
\usepackage{graphicx}
\usepackage{mdwlist}
\usepackage{multirow}
\usepackage{amsmath}
\usepackage{enumitem} 
\usepackage{calc}
\usepackage{colortbl}
\usepackage[ruled,linesnumbered]{algorithm2e}
\usepackage{booktabs}
\usepackage{mathtools}
\usepackage[export]{adjustbox}
\usepackage{balance}
\newcommand{\para}[1]{\paragraph{\textnormal{\textbf{#1}}}}
\renewcommand{\vec}[1]{\mathbf{#1}}

\newcommand{\matdeclare}[3]{\vec{#1} \in \mathbb{R}^{#2\times #3}}
\renewcommand{\th}[1]{#1{^\mathrm{th}}}
\newcommand{\pr}{Pearson's-$\rho$}
\newcommand{\wsup}{WS-NeurQPP}
\newcommand{\ours}{Deep-QPP}

\DeclareMathOperator{\sigmoid}{Sigmoid}

\DeclareMathOperator{\maxpool}{MaxPooling}
\DeclareMathOperator{\avgpool}{AvgPooling}
\DeclareMathOperator{\sgn}{sgn}

\theoremstyle{definition}

\newtheorem{example}{Example}[section]

\AtBeginDocument{%
  \providecommand\BibTeX{{%
    \normalfont B\kern-0.5em{\scshape i\kern-0.25em b}\kern-0.8em\TeX}}}

\copyrightyear{2022}
\acmYear{2022}
\setcopyright{acmcopyright}
\acmConference[WSDM '22]{Proceedings of the Fifteenth ACM International Conference on Web Search and Data Mining}{February 21--25, 2022}{Tempe, AZ, USA}
\acmBooktitle{Proceedings of the Fifteenth ACM International Conference on Web Search and Data Mining (WSDM '22), February 21--25, 2022, Tempe, AZ, USA}
\acmPrice{15.00}
\acmDOI{10.1145/3488560.3498491}
\acmISBN{978-1-4503-9132-0/22/02}

\begin{document}

\fancyhead{}

\title{\ours: A Pairwise Interaction-based Deep Learning Model for Supervised Query Performance Prediction}

\author{Suchana Datta}
\affiliation{%
  \institution{University College Dublin, Ireland}
  \country{}
}
\email{suchana.datta@ucdconnect.ie}
\author{Debasis Ganguly}
\affiliation{%
  \institution{University of Glasgow, UK}
  \country{}
}
\email{debasis.ganguly@glasgow.ac.uk}
\author{Derek Greene}
\affiliation{%
  \institution{University College Dublin, Ireland}
  \country{}
}
\email{derek.greene@ucd.ie}
\author{Mandar Mitra}
\affiliation{%
  \institution{Indian Statistical Institute, India}
  \country{}
}
\email{mandar@isical.ac.in}

\begin{abstract}
Motivated by the recent success of end-to-end deep neural models for ranking tasks, we present here a supervised end-to-end neural approach for query performance prediction (QPP).
In contrast to unsupervised approaches that rely on various statistics of document score distributions, 
our approach is entirely data-driven.
Further, in contrast to weakly supervised approaches, our method also does not rely on the outputs from different QPP estimators.
In particular, our model leverages information from the semantic interactions between the terms of a query and those in the top-documents retrieved with it. The architecture of the model comprises
multiple layers of 2D convolution filters
followed by a feed-forward layer of parameters.
Experiments on standard test collections demonstrate that our proposed supervised approach outperforms other state-of-the-art supervised and unsupervised approaches.   
\end{abstract}




\maketitle

\section{Introduction} \label{sec:intro}
The evaluation of information retrieval systems is a challenging problem to solve outside the realm of the Cranfield paradigm \cite{cranfield}, i.e., in situations when there are no relevance assessments available, such as those in deployed search systems used by real-life users beyond the laboratory environment.
Query performance prediction (QPP) \cite{croft_qpp_sigir02, carmel_qpp_sigir06, wig_croft_SIGIR07, kurland_tois12, query_variants_kurland}, therefore,
remains an important and active area of research, because of its usefulness in estimating the quality of a retrieval system on a wide range of queries.
%
%
The output of a QPP estimator function $\phi(Q)$ is a likelihood score ($\in \mathbb{R}$), which given a query $Q$, predicts the retrieval quality of the query. It may therefore, in a sense, be considered to represent how \emph{easy} (or specific) the query is, because the higher the predicted estimate, the higher is the likelihood that a retrieval model will perform well for the query.

The majority of existing QPP methods relies on devising a suitable heuristic function for predicting the likelihood of how easy a query will be for a retrieval system. 
Typically, this is estimated by computing the probability of how specific or well-formulated the query is.
The specificity measures are computed using either: i) an aggregate
of collection statistics over query terms
commonly known as \emph{pre-retrieval} QPP estimators \cite{hauff_2010,pre-ret_survey_cikm08}; or by ii) leveraging information from the top-retrieved documents, e.g., assessing the skewness of document similarity scores \cite{wig_croft_SIGIR07,kurland_tois12}, or measuring the topical differences between the set of top-retrieved documents and the rest of the collection \cite{croft_qpp_sigir02}.

Supervised deep neural ranking models have recently been shown to improve retrieval effectiveness over their unsupervised statistical counterparts \cite{drmm,knrm_SIGIR17,conv_knrm,neural_ranking_weak_sup_zamani_sig-17,colbert_sigir20}. 
In contrast to preset similarity functions (e.g. BM25 or LM),
these supervised models rely on data-driven parametric learning of similarity functions, usually leveraging an interaction mechanism between the similarities of the embedded representations of constituent words of queries and their retrieved documents \cite{drmm,knrm_SIGIR17,conv_knrm}.

While the benefits of using supervised approaches have predominantly been established for ranking \cite{jimmy_lin_web_rank_func_sig-11,fast_learning_to_rank_sig-18,colbert_sigir20,neural_ranking_weak_sup_zamani_sig-17}
and recommendation tasks \cite{neural_reco_recsys-19,narm_cikm-17,diffusion_model_sig-19,context_seq_model_dlrs-17}, 
the exploration of supervised approaches for QPP has been limited. 
The only supervised QPP approach, that we are aware of to the best of our knowledge at the time of writing this paper, is the study \cite{hamed_neuralqpp} which used a combination of features (such as retrieval scores) and word embedded vectors 
to learn an optimal way of \emph{combining} a number of different QPP estimates into a single one, thereby outperforming
the effectiveness achieved by each individually.
A major limitation of \cite{hamed_neuralqpp} is that the training procedure involves weak supervision over a number of estimators to find an optimal combination. In contrast, our proposed method is solely data-driven because it does not rely on other estimators. Moreover, our method is strictly supervised as opposed to the weak supervision employed in \cite{hamed_neuralqpp}.

\subsubsection*{\emph{\textbf{Contributions}}}
In summary, the key contributions of this paper include --
\begin{enumerate}
[leftmargin=*]
\item
\textbf{An end-to-end supervised QPP model}, where instead of learning to optimize the relative importance of different predictors \cite{hamed_neuralqpp}, our model learns a comparison function of relative specificity (estimated retrieval quality) between query pairs.
\item
\textbf{Early interactions between query-document pairs}, where similar to the deep relevance matching model (DRMM) \cite{drmm}, our model makes use of early interactions between a query and its top-retrieved set of documents. We argue that this way of constituting the input improves the model's capacity to generalize better as opposed to the late interaction between the content of the queries and the documents \cite{hamed_neuralqpp}.  
\end{enumerate}

\begin{figure}[t]
\centering
\includegraphics[width=\columnwidth]{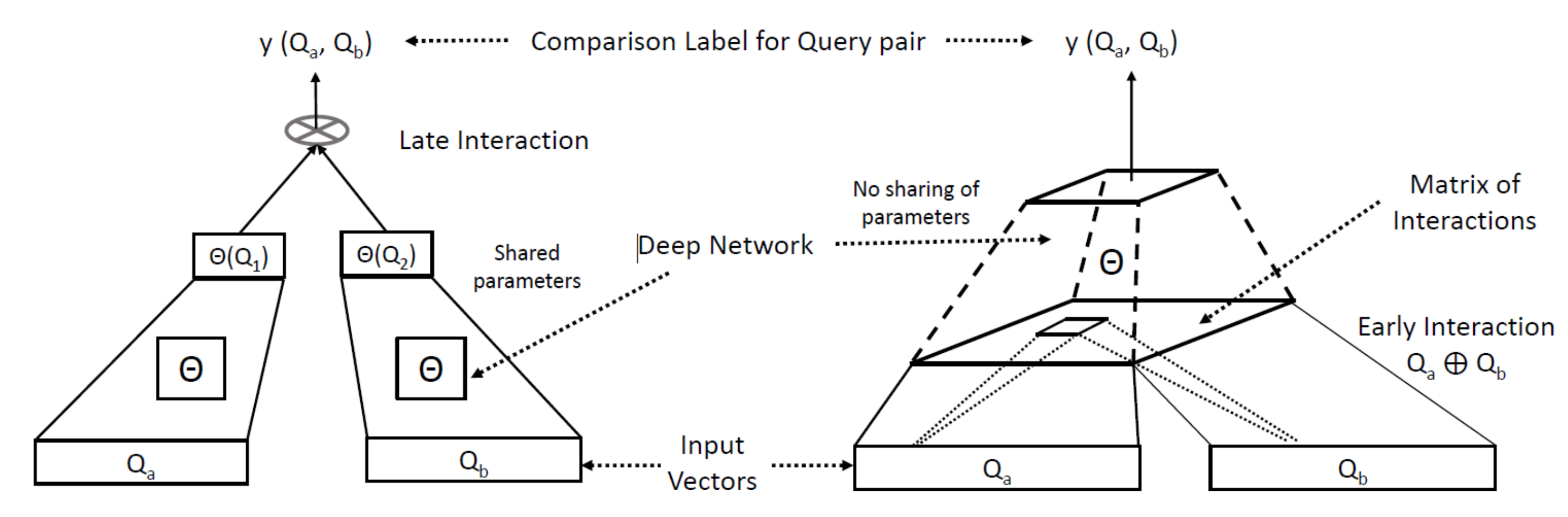}
\caption{
\small
While representation-based models rely on \emph{late interaction} involving shared parameters (left), interaction-based models, on the other hand, make use of \emph{early interactions} transforming paired instances into a single input.
}
\label{fig:latevsearly}
\end{figure}

\section{\ours~Model Description} \label{sec:modeloverview}
We first describe the working principle of our approach which is based on capturing term-semantics interaction at two levels, first,
at the \emph{intra-query} level of modeling, the interaction between the queries themselves and their top-retrieved documents,
and then at the
\emph{inter-query} level, to model their relative specificity measures.


\subsection{Representation vs. Interaction}
A fundamental difference between a representation-based model and an interaction-based model \cite{drmm} is illustrated in Figure \ref{fig:latevsearly}. The former first constructs a representation of each instance from a pair of inputs, and then optimizes this representation so as to maximize the likelihood of predicting a function involving this pair (as seen from the left diagram of Figure \ref{fig:latevsearly}). In contrast, an interaction-based model first \emph{transforms} a paired data instance into a single instance via an interaction operator $\oplus: \mathbb{R}^d \times \mathbb{R}^d \mapsto \mathbb{R}^p$, where $d$ and $p$ are the sizes of the raw and the transformed inputs, respectively.

We now discuss the type of interaction suitable for a supervised deep QPP approach.
For QPP, the objective function that should be learned from the reference labels is a comparison between a pair of queries, $Q_a$ and $Q_b$. More concretely, this comparison is an indicator of the relative difficulty between the queries,
i.e., whether $Q_a$ is more difficult than $Q_b$ or vice versa.

\begin{figure}[t]
\centering
\includegraphics[width=.9\columnwidth]{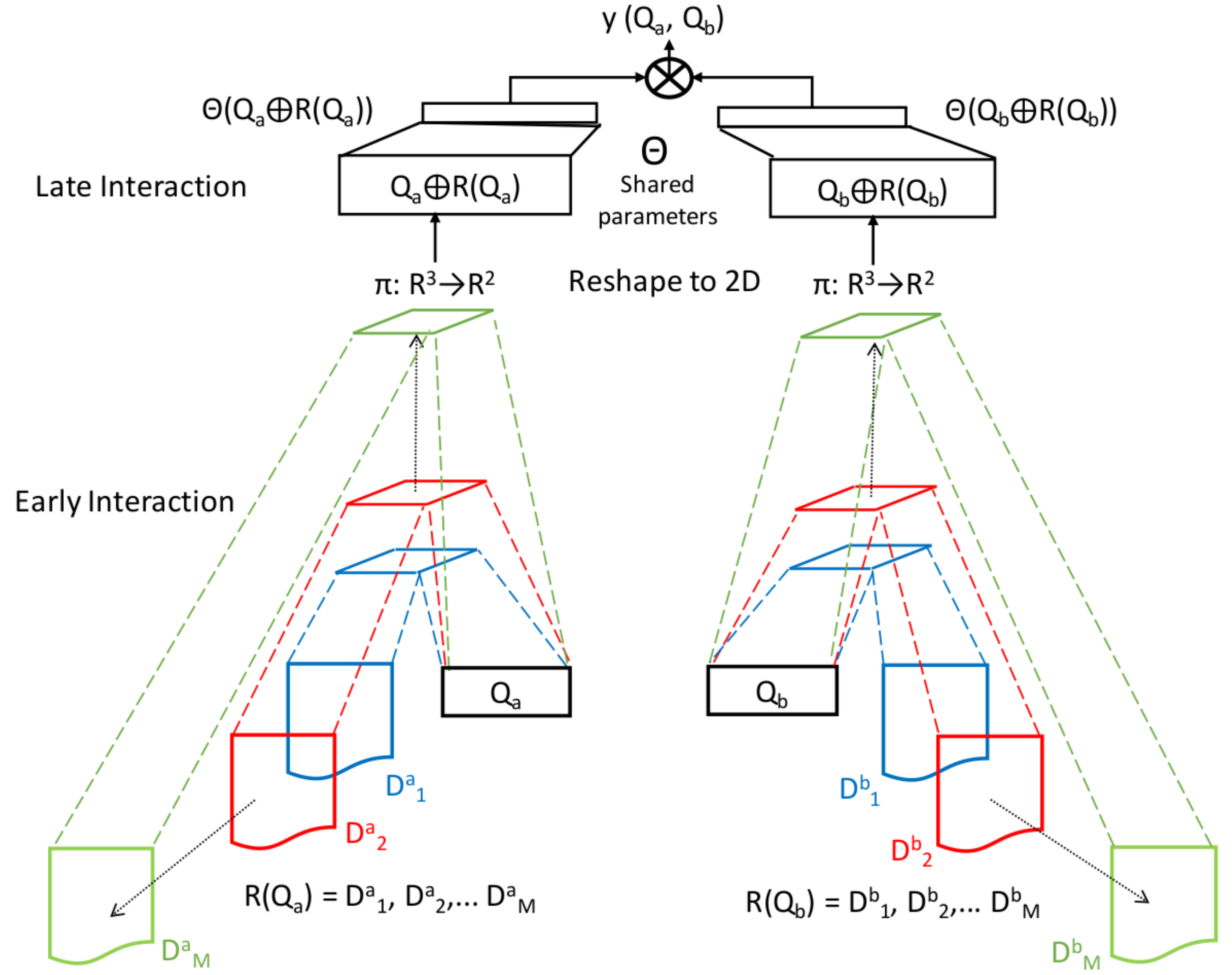}
\caption{
\small
Unlike an entirely representation-based or interaction-based model (Figure \ref{fig:latevsearly}), our model combines the benefits of both early and late interactions, to address: a) the interaction of the terms in the top-retrieved documents of a query with the constituent terms of the query itself; b) the characteristic pattern of these interactions to estimate the comparison function $y(Q_a, Q_b)$ between a pair of queries. Each individual query-document interaction is shown with a different color. 
}
\label{fig:hybrid}
\end{figure}

While pre-retrieval QPP approaches only rely on the information
from a query itself (e.g., aggregate collection statistics for its terms \cite{hauff_2010,pre-ret_survey_cikm08}), it has been shown that post-retrieval approaches, which make use of additional information from the top-retrieved documents of a query \cite{wig_croft_SIGIR07,kurland_tois12}, usually perform better. Motivated by this,
we also include information from the top-retrieved documents in the form of \emph{early interactions} (which we refer to as the \emph{intra-query} interactions).
The parameters of these interactions are then optimized with the help of a \emph{late interaction} between the queries, which seeks to capture the important characteristic differences of these early interactions towards identifying which query among the pair is easier.
An overview of our model is shown in Figure \ref{fig:hybrid}.

\subsection{Query-Document Interactions} \label{ss:qd-interaction}

In unsupervised post-retrieval QPP approaches, the interaction between the terms in a query and those of the top-retrieved set takes the form of statically defined functions, which aim to capture how distinct the top-retrieved set is with respect to the collection (e.g., NQC \cite{kurland_tois12} uses the skewness of document retrieval scores, while WIG \cite{wig_croft_SIGIR07} measures the information gain from the top-retrieved set with respect to the collection).
The intra-query interaction shown in Figure \ref{fig:hybrid} involves computing an interaction between the terms of a query and those in its top-retrieved set of documents. This interaction then acts as an input source to learn an optimal specificity function automatically from the data.

\para{Documents to consider for interaction}
A common principle that works well as a specificity estimator for post-retrieval QPP approaches, 
is to measure the distinctiveness between the set of documents towards the top-ranks from the rest of the retrieved set. The standard deviation of the document similarity scores in NQC (i.e., expected difference from the average score), acts as an estimate for the topic distinctiveness of the top set. 

Motivated by this insight, in our approach, instead of using only a set of top-$k$ documents, we use information from both the upper and the lower parts of a ranked list. The objective is to capture the differences in the interaction patterns of a set of highly similar (the upper part of a ranked list) and not-so-highly similar documents (the lower part) as useful cues for QPP.

As notations, we denote the set of documents considered for interaction with a query $Q$ as $R(Q)$, which is comprised of a total of $M=t+b$ documents, including the top-$t$ and the bottom-$b$ ranked ones. The index of the bottom-most document considered for interaction computation is specified by a parameter $N$. This means that the lower part of the ranked list, comprised of $b$ documents are, in fact, those ranked from $N$ to $N-b+1$.
For example, a value of $t=10$ and $b=20$ means that $R(Q) = \{D_1,\ldots,D_{10}\} \cup \{D_{81},\ldots,D_{100}\}$.

In our experiments, we treat $t$ and $b$ as hyper-parameters (see Section \ref{ss:results}), and restrict $N$ to a value of $100$ because it is unlikely that any evidence from documents beyond the top-$100$ would be useful for the QPP task.


\para{Interaction between each query term and a document}

We now describe how we compute the query-document interaction matrices for each document $D\in R(Q)$ for a query $Q$.
As a first step, we calculate the cosine similarities between the embedded representations of terms -- one from the query $Q_a$ and the other from the document $D^a_i$. Similar to \cite{drmm}, the distribution of similarities between the $\th{j}$ query term $q_j$ and constituent terms of $D^a_i$ is then transformed into a vector of fixed length $p$ by the means of computing a histogram of the similarity values over a partition of $p$ equi-spaced intervals defined over the range of these values (i.e. the interval $[-1, 1)$). 
The $\th{\beta}$ component ($\beta=1,\ldots,p$) of this interaction vector is given by the count of how many terms yield similarities that lie within the $\th{\beta}$ partition of $[-1, 1)$, i.e., 
\begin{equation}
(q_j\oplus D^a_i)_\beta =
\sum_{w \in D^a_i} \mathbb{I}\big[\frac{2(\beta-1)}{p}-1 \leq
\frac{\vec{q}_j \cdot \vec{w}}{|\vec{q}_j| |\vec{w}|}
< \frac{2\beta}{p}-1\big] \label{eq:interaction-qterm-doc-noidf}, 
\end{equation}
where both $\vec{q}_j \in \mathbb{R}^d$ and $\vec{w} \in \mathbb{R}^d$, and the interaction vector $q_i\oplus D^a_i \in \mathbb{R}^p$,
and $\mathbb{I}[X] \in \{0, 1\}$ is an indicator variable which takes the value of $1$, if a property $X$ is true and $0$ otherwise.

\begin{example}\label{example:binning}
If $p=4$, the interval $[-1, 1)$ is partitioned into the set $\{[-1, -0.5), [-0.5, 0), [0, 0.5), [0.5, 1)\}$. For a 3-term document $d$, if the cosine similarities are $0.2$, $-0.3$ and $0.4$ with respect to a query term $q$, then $q \oplus d = (0, 1, 2, 0)$. 
\end{example}

\para{Collection statistics based relative weighting}
The specificity (i.e., collection statistics, such as idf) of query terms contributes to the effective estimate of QPP scores both in pre-retrieval and post-retrieval approaches. 
We, therefore, incorporate the idf values of each query term as a factor within the interaction patterns to relatively weigh the contributions from the interaction vectors $q_j \oplus D^a_i$. In our proposed approach, we use a generalized version of Equation \ref{eq:interaction-qterm-doc-noidf}, where we incorporate the idf factor as a part of the interaction vector components, i.e.,
\begin{equation}
(q_j\oplus D^a_i)_\beta = 
\log(\frac{N_0}{n(q_j)})
\sum_{w \in D^a_i} \mathbb{I}\big[\frac{2(\beta-1)}{p}-1 \leq
\frac{\vec{q}_j \cdot \vec{w}}{|\vec{q}_j| |\vec{w}|}
< \frac{2\beta}{p}-1\big] \label{eq:interaction-qterm-doc}, 
\end{equation}
where
$n(q_j)$ denotes the number of documents in the collection where the $\th{j}$ query term $q_j$ occurs, and $N_0$ denotes the total number of documents in the collection.

\para{Overall interaction between a query and a document}
Each $p$-dimensional interaction vector computed for the $\th{j}$ query term forms the $\th{j}$ row of the overall interaction matrix between the query $Q_a$
and the $\th{i}$ document $D^a_i$. The overall interaction matrix, $Q_a\oplus D^a_i \in \mathbb{R}^{k\times b}$ is thus given by
\begin{equation}
Q_a\oplus D^a_i = 
[(q_1\oplus D^a_i)^T, \ldots, (q_k\oplus D^a_i)^T]^T \label{eqn:interaction-q-doc},
\end{equation}
where $k$ is a preset upper limit of the number of terms in a query. A zero-padding is used for the row indices exceeding the number of query terms, i.e., $(q_j\oplus D^a_i)=\{0\}^b,\, \forall j>|Q_a|$.
Referring back to Figure \ref{fig:hybrid}, each $k\times p$ interaction matrix between a query $Q_a$ and a document $D^a_i$ corresponds to a colored rectangle (shown in the planes above the queries and documents).


\begin{figure*}[t]
\centering
\includegraphics[width=.9\textwidth]{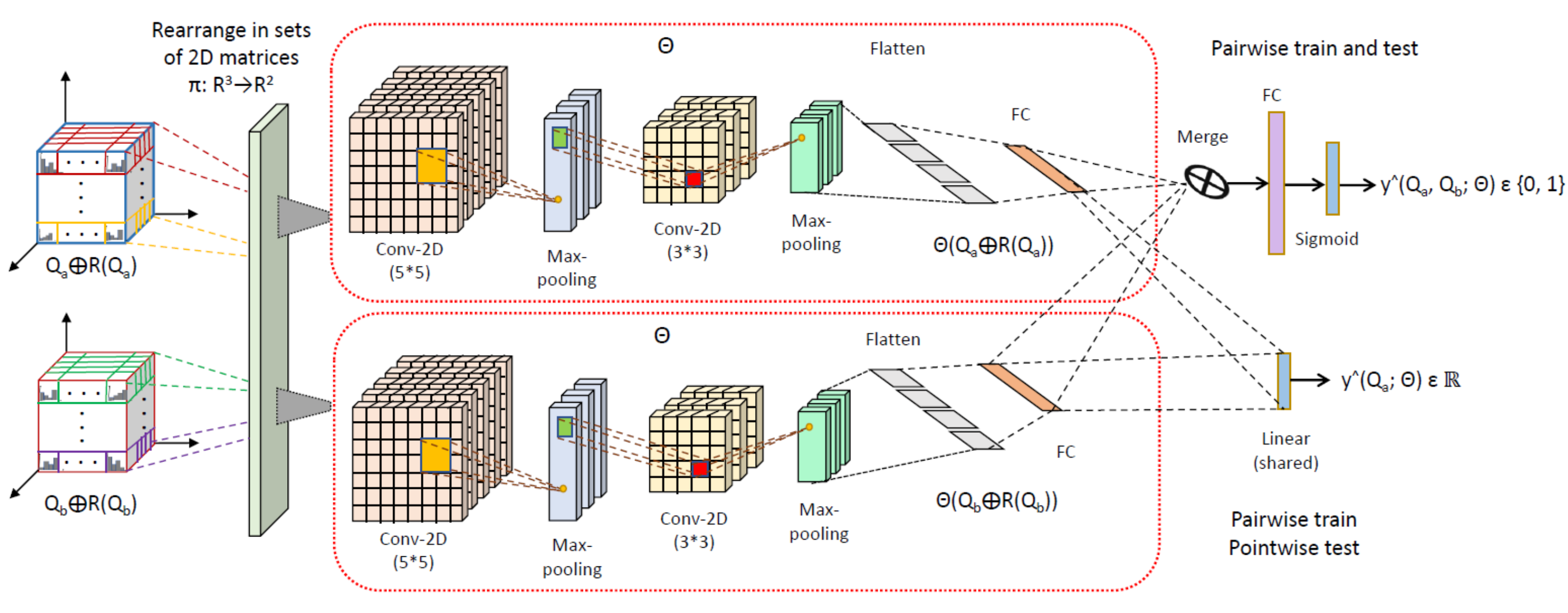}
\caption{
\small
Our proposed end-to-end QPP model comprising a Siamese network of shared parameters of layered convolutional feature extraction, followed by either i) merge (concatenation) and a fully connected (FC) layer with a $\sigmoid$ loss for pairwise testing (Equation \ref{eq:pairwise-test-loss}) yielding a binary comparison indicator between a pair, or ii) a linear activation layer with pairwise hinge loss for pointwise testing yielding a score for a given query (Equation \ref{eq:pointwise-test-loss}). Since the interaction for MDMQ and SDSQ are matrices with a single row only, the two layers of convolution filter sizes for these approaches are $1\times 5$ and $1\times 3$ (see Section \ref{ss:reshape}).
}
\label{fig:overview}
\end{figure*}

\para{Interaction between a query and its top-retrieved set}
Finally, each individual document-query interaction matrix, when stacked up one above the other in the order of the document ranks, yields an interaction tensor of order $M\times k\times p$. Formally,
\begin{equation}
Q_a\oplus R(Q_a) =
\begin{bmatrix}
Q_a\oplus D^a_1\\
\ldots\\
\ldots\\
Q_a\oplus D^a_M
\end{bmatrix}
\label{eq:interaction-tensor}
\end{equation}


\subsection{Layered Convolutions for QPP}
After constructing the local interactions of a query with its top-retrieved set of documents, i.e. the intra-query interactions, the next step is to extract convolutional features from the $3^{rd}$ order interaction tensor, $Q_a\oplus R(Q_a) \in \mathbb{R}^{M\times k \times b}$ between a query $Q_a$ and its top-retrieved set $R(Q_a)$. To this end, we first need to slice the $3^{rd}$ order tensor into separate matrices ($2^{nd}$ order tensors), on each of which, 2D convolution can be applied to extract distinguishing features from the raw data of query-document interactions.
Before describing the 
ways to
slicing the tensor into matrices in Section \ref{ss:reshape}, we first briefly describe the 
architecture that we employ to extract useful features from the lower-dimensional slices of the interaction tensor.

\para{Brief background on 2D convolution}

We do not explain the background of 2D convolution operation \cite{vcortex} in detail. Formally speaking, 
if $\matdeclare{X}{M}{P}$ represents an input data matrix, and if $\matdeclare{W^{(\mathit{l})}}{k_l}{k_l}$ ($k_l\mod 2 = 1$, i.e., $k_l$ an odd number) denotes the kernel weight matrix of the $\th{l}$ layer,
conveniently represented as
$(W^{(l)}_{-\lfloor k/2\rfloor},\ldots,0,\ldots,W^{(l)}_{\lfloor k/2\rfloor})$,
then the outputs of layer-wise convolution, generally speaking, are given by 
\begin{equation}
\vec{h}^{(l)}_{r,c} = f^{(l)}(
\sum_{i=-\lfloor k/2\rfloor}^{\lfloor k/2\rfloor}
\sum_{j=-\lfloor k/2\rfloor}^{\lfloor k/2\rfloor}
\vec{W}^{(l)}_{i,j}\vec{h}^{(l-1)}_{r+i,c+j}
),
\label{eq:convlayers}
\end{equation}
for each $l=1,\ldots,L$ ($L$ being the total number of layers),
where $\vec{h}^{(l-1)}\! \in \!\mathbb{R}^{M^{(k-1)} \times P^{(k-1)}}$ 
is the output obtained from the previous layer of the convolution filter, with $h^{(1)}\!=\!X$, $M^{(1)}\!=\!M$ and $P^{(1)}=P$. The function $f^{(l)}$ is an aggregation function that, generally speaking, progressively reduces the size of the convolutional representations, $\vec{h}^{(l)}$, across layers. Aggregation methods commonly applied in computer vision include the $\maxpool$ \cite{christlein2019deep,maxpool_nips-15} and $\avgpool$ \cite{pool_sig-14} functions.


\para{Late interactions with convolutional features}
A more detailed view of the late interaction across a query pair is shown in Figure \ref{fig:overview}. Referring to the notation from Equation \ref{eq:convlayers}, we employ $L=2$ (i.e. use a total of 2 convolution layers), and use $k_1=5$ and $k_2=3$ (i.e. a 5x5 filter for the first layer and a 3x3 for the second one). The aggregate function, $f^{(l)}$, of each layer $l$ is set to the $\maxpool$ operation.

After extracting the convolutional features for each query vs. top-documents interaction tensor (shown as the two cuboids towards the extreme left of Figure \ref{fig:overview}), we employ the standard practice of merging the convolutional filter outputs of each query into a single vector (shown as the `merge' operation) \cite{merge-cnn_20, merge_cnn_corr_20}.
Following the merge operation, which now combines abstract features extracted from the local interactions of the two queries into a single vector, we apply a fully connected dense layer.
Depending on whether we test the network in a pointwise or pairwise manner, the loss function is set to either the $\sigmoid$ function or a function that seeks to maximize the accuracy of the comparison function between pairs. Section \ref{ss:training} provides more details on the network training process. 

\subsection{Reshaping the Interaction Tensor} \label{ss:reshape}
There exists a number of different choices for slicing up the interaction tensor of Equation \ref{eq:interaction-tensor} into a set of matrices for the purpose of separately applying 2D convolution on each and then combining the features, shown as the \emph{reshaping} function $\pi:\mathbb{R}^3 \mapsto \mathbb{R}^2$ in Figure \ref{fig:overview}. 
We now discuss each alternative and examine their pros and cons in the context of the QPP problem.

As our nomenclature, we characterize reshaping functions by whether the information across i) top-retrieved documents are merged together, or across ii) query-terms are merged together. A part of the name thus uses the characters \texttt{D} to denote the top-retrieved set, and \texttt{Q} to denote query terms. To indicate `merging', we use the letter `\texttt{M}' and to denote its counterpart, we use the letter `\texttt{S}' (separate). For instance, the name \texttt{MDMQ} means that the information from both top-documents and query terms are merged together.

\para{MDMQ (Merged Documents Merged Query-terms)} This is the most coarse-grained way to reduce the dimensionality of the interaction tensor of order $3$ (Equation \ref{eq:interaction-tensor}) by reducing the $M\times k \times p$ tensor to a flattened vector of dimensionality $Mkp$, which can still be imagined to be a matrix of dimension $1\times Mkp$ allowing 1D convolutions to be applied. This method extracts abstract features at an aggregate level rather than for individual documents separately. This may not be desirable because, in standard QPP methods such as WIG and NQC, an individual contribution from each document score is responsible for the predicted specificity measure.

\para{SDMQ (Separate Documents Merged Query-terms)} This corresponds to the most natural way of grouping an interaction tensor, $Q\oplus R(Q)$, by considering the $\th{i}$ row for each $i=1,\ldots,M$, $Q\oplus D_i$, as a matrix of dimension $k\times p$. This method allows the extraction of abstract features from each document separately in relation to the whole query. Thus, it takes into account the compositionality of the query terms, and at the same time avoids mixing information across documents. This conforms to how most unsupervised post-retrieval QPP methods actually work. 

\para{MDSQ (Merged Documents Separate Query-terms)} Contrary to grouping the interaction tensor row-wise, for this method we slice out the constituent matrices column-wise. Each matrix is thus of dimension $M\times p$, and there are a total $k$ of them, on each of which we apply 2D convolution for feature extraction. This QPP method thus does not take into account the compositionality of the constituent query terms while considering the semantic interactions. Rather it treats the whole set of top-retrieved documents in an aggregated manner, which is also somewhat counter-intuitive because a document at the very top rank should be treated in a different manner from the very bottom one, i.e. the one at $\th{M}$ rank.

\para{SDSQ (Separate Documents Separate Query-terms)}
This is the most fine-grained approach, which considers every interaction vector between the $\th{j}$ query term and $\th{i}$ document (see Equation \ref{eq:interaction-qterm-doc} as a separate candidate for convolutional feature extraction. Each such interaction vector between a query-term and a document is of dimension $p$ and there are a total of $Mk$ such vectors. As with the MDMQ approach, we apply 1D convolution on these vectors.

%
A point to note is that, although Figure \ref{fig:overview} shows the convolution filters as $5\times 5$ and $3\times 3$, for MDMQ and SDSQ approaches, these filters are of size $1\times 5$ and $1\times 3$ respectively.


\section{\ours~Training} \label{ss:training}

The network in Figure \ref{fig:overview} is trained with instances of query pairs with two different objectives -- pointwise and pairwise.
In the pairwise case, the network directly learns the comparison function, i.e. a binary indicator of the anti-symmetric relation between a query pair. On the other hand, the pointwise objective aims to predict a QPP score, instead of the relative order of specificity between a pair. Before describing the objectives, we first provide details on obtaining the data instances and the reference labels.

\subsection{Instances and Ground-truth Labels}
Given a training set of queries $\mathcal{Q}=\{Q_1,\ldots,Q_m\}$, we construct the set of all unordered pairs of the form $(Q_a, Q_b)$, where $\forall a,b \leq m$ and $b > a$.
The reference label, $y(Q_i, Q_j)$, of a paired instance is determined by a relative comparison of the retrieval effectiveness obtained by a system with a target metric.
The retrieval effectiveness, in turn, is computed with the help of available relevance assessments.
Formally speaking, if
$\mathcal{M}$ denotes an IR evaluation measure (e.g., average precision or AP), which is a function of i) the known set of relevant documents - $\mathcal{R}(Q)$ for a query $Q \in \mathcal{Q}$, and ii) the set of documents retrieved with a model $\mathcal{A}$ (e.g., LM-Dir \cite{lmdir}), then
\begin{equation}
y(Q_a, Q_b) =
\sgn(
\mathcal{M}(Q_a; \mathcal{R}(Q_a)) -
\mathcal{M}(Q_b; \mathcal{R}(Q_b))
) \label{eq:reflabels},
\end{equation}
where $\sgn(x)=0$ if $x\leq 0$ or $1$ otherwise.
%


For all our experiments, we used AP@100 and nDCG@20 as the target metric $\mathcal{M}$. As the IR model, $\mathcal{A}$, we employ
LM-Dir with the smoothing parameter $\mu = 1000$ following QPP literature \cite{kurland_tois12}. We emphasize that the results of our experiments are mostly insensitive to the choice of either the target metric used or the IR model employed.


\subsection{Pairwise Objective} \label{ss:pairwise}
For the pairwise objective, the \ours~model is trained to maximize the likelihood of correctly predicting the indicator value of the comparison between a given pair of queries. The purpose here is to learn a data-driven generalization of the comparison function. During the testing phase, the model outputs a predicted value of the comparison between a pair of queries unseen during the training phase. 
The output layer for the pairwise objective thus constitutes a $\sigmoid$ layer, predicting values of $y(Q_a, Q_b)$ (see Equation \ref{eq:reflabels}) as a function of the network parameters denoted as $\hat{y}(Q_a, Q_b;\Theta)$.
During the training phase, the parameter updates seek to minimize the standard square loss
\begin{equation}
\mathcal{L}(Q_a, Q_b) = (y(Q_a,Q_b)-\hat{y}(Q_a, Q_b;\Theta))^2
\label{eq:pairwise-test-loss}
\end{equation}
between the ground-truth and the predicted labels.

\subsection{Pointwise Objective}

For pointwise testing, as a test input, the network takes a single query $Q$, as opposed to the pair of queries in the pairwise situation from Section \ref{ss:pairwise}. Instead of predicting a binary indicator comparison, the network predicts a score $\hat{y}(Q;\Theta)$ that can be used as an estimated measure of specificity of $Q$.
To allow for pointwise testing, the output from the shared layer of parameters goes into a linear activation unit predicting a real-valued score $\hat{y}(Q;\Theta)$, which is a function of one query (rather than a pair), as can be seen from the bottom-right part of the Figure \ref{fig:overview}. Rather than training the network on a merged representation of a query pair, the loss function includes separate contributions from the two parts of the network corresponding to each query, the objective being to update the parameters for maximizing the comparison agreements between the reference and the predicted scores. Specifically, we minimize the following hinge loss:
%
\begin{equation}
\mathcal{L}(Q_a,Q_b) = \max (0, 1-
\sgn(
y(Q_a, Q_b)\cdot
(\hat{y}(Q_a;\Theta) -\hat{y}(Q_b;\Theta))
)
).
\label{eq:pointwise-test-loss}
\end{equation}

\section{Experiments} 
\label{sec:setup}

\subsection{Datasets and Hyper-parameters} \label{ss:datasets}
\begin{table}[t]
\small
\centering
\begin{tabular}{@{}lccccc@{}}
\toprule
Coll & \#Docs & Topic Set& $|\mathcal{Q}|$ & Avg. Q\_len & Avg. \#Rel \\
\midrule
Disks 4 \& 5 & 528,155 & TREC-Rb & 249 &2.68 & 71.21 \\
CWeb09B-S70 & 29,038,220 & TREC-Web & 200 & 2.42 & 16.03 \\
\bottomrule
\end{tabular}
\caption{
\small
Dataset Characteristics
(the suffix `S70' indicates that documents detected as spams with confidence scores higher than $70\%$ were removed from the collection).
}
\label{tab:datastats}
\end{table}

\para{Collections}
We experiment with two standard ad-hoc IR test collections, namely the TREC Robust (comprised of news articles) and the ClueWeb09B \cite{web_track_10} (comprised of crawled web pages). For the ClueWeb experiments, we used the Waterloo spam scores 
\cite{waterloospam}
to remove documents which were detected to be spam with confidence $> 70\%$. We denote this subset as CWeb09B-S70 in Table \ref{tab:datastats}.

\para{Train and test splits}
Since our proposed \ours~method is a supervised one, the method first requires a training set of queries to learn the model parameters and then a test set for evaluating the effectiveness of the model. 
Following the standard convention in the literature, e.g. \cite{hamed_neuralqpp,kurland_tois12,query_variants_kurland}, we employ repeated partitioning (specifically, 30 times) of the set of queries into 50:50 splits and report the average values of the correlation metrics (see Section \ref{ss:evalmetrics}) computed over the 30 splits.

A major difference of our setup compared to existing QPP approaches is the use of the training set. While the training set for unsupervised approaches serve the purpose of \emph{tuning the hyper-parameters} of a model by grid search, in our case, it involves \emph{updating the learnable parameters} of the neural model by methods such as stochastic gradient descent. 

\para{Hyper-parameter tuning} The most common hyper-parameter for existing unsupervised QPP approaches is the number of top-$M$ documents considered for computing the statistics on the document retrieval scores, as in NQC and WIG, or to estimate a relevance feedback model, as in Clarity and UEF (see Section \ref{ss:baselines} for more details). We tune this parameter via grid search on the training partition.
As prescribed in \cite{hamed_neuralqpp},
the values used in grid search were $\{$5, 10, 15, 20, 25, 50, 100, 300, 500, 1000$\}$.

\subsection{Baselines} \label{ss:baselines}
We compare our supervised 
\ours~approach with a number of standard unsupervised QPP approaches, and also a more recent weak supervision-based neural approach \cite{hamed_neuralqpp}. In our investigation, we do not include QPP methods that leverage external information, such as query variants \cite{kurland_qpp_rlm}. Using query variants has been shown to improve the effectiveness of unsupervised QPP estimators and it is also likely that including them in our supervised end-to-end approach will may also lead to further improvement in its performance. However, since the main objective of our experiments is to investigate if a deep QPP model can outperform existing ones, we leave the use of external data for future exploration.
Moreover, we also do not include the pre-retrieval QPP approaches, such as avg. idf etc., because they have been reported to be outperformed by post-retrieval approaches in a number of existing studies \cite{croft_qpp_sigir02,wig_croft_SIGIR07,kurland_tois12,hamed_neuralqpp}. 

\subsubsection{Unsupervised Approaches}
This refers to existing methods that make use of term weight heuristics to measure the specificity estimates of queries. The underlying common principle on which all these methods rely is the assumption that, if the set of top-documents retrieved for a query is substantially different from the rest of the collection, then the query is likely to be indicative of unambiguous information need. This makes it a potentially good candidate for achieving effective retrieval results. These methods mainly differ in the  way in which they calculate the similarity of the top-retrieved set of documents from the rest of the collection. 

\para{Clarity \cite{croft_qpp_sigir02}}
This method estimates a relevance model (RLM) \cite{Lavrenko_RLM2001:RBL:383952.383972} distribution of term weights from a set of top-ranked documents, and then computes its KL divergence with the collection model - the higher the KL divergence (a distance measure) the higher is the query specificity.

\para{WIG \cite{wig_croft_SIGIR07}}
As its specificity measure, weighted information gain (WIG) uses the aggregated value of the information gain with each document (with respect to the collection) in the top-retrieved set. The more topically distinct a document is from the collection, the higher its gain will be. Hence, the average of these gains characterizes how topically distinct is the overall set of top-documents.

\para{NQC \cite{kurland_tois12}}
Normalized query commitment (NQC) estimates the specificity of a query as the standard deviation of the RSV's of the top-retrieved documents with the assumption that a lower deviation from the average (indicative of a flat distribution of scores) is likely to represent a situation where the documents at the very top ranks are significantly different from the rest. NQC thus makes use of not only the relative gain of a document score from the collection (similar to WIG) but also the gain in a document's score with respect to the average score.

\para{UEF \cite{uef_kurland_sigir10}}
The UEF method assumes that information from some top-retrieved set of documents are more reliable than others. As a first step, the UEF method estimates how robust is a set of top-retrieved documents by checking the relative stability in the rank order before and after relevance feedback (by RLM). The higher the perturbation of a ranked list post-feedback for a query, the greater is the likelihood that the retrieval effectiveness of the initial list was poor, which in turn suggests that a smaller confidence should be associated with the QPP estimate of such a query.

\subsubsection{Supervised Approaches}
Our choice of supervised baselines is guided by two objectives - \emph{first}, to show that (strong) supervision using the ground-truth of relative query performance is better than the existing approach of weak supervision on QPP estimation functions \cite{hamed_neuralqpp}, and \emph{second}, to demonstrate that a mixture of both early and late interactions (i.e., a hybrid of both content and interaction-focused approaches) is better than purely content-based ones (see Figures \ref{fig:latevsearly} and \ref{fig:hybrid}).
%
\para{Weakly Supervised Neural QPP (WS-NeurQPP) \cite{hamed_neuralqpp}}
The main difference between WS-NeurQPP and Deep-QPP lies in the source of information used and also the objective of the neural end-to-end models. WS-NeurQPP uses weak supervision to approximate the scores of individual QPP estimators so as to learn an optimal combination.
As inputs, it uses the retrieval scores, along with the word embedded vectors. However, in contrast to our approach, it does not use interactions between terms and is hence a purely representation-based approach. 

\para{Siamese Network (SN)} 
This approach is an ablation of the Deep-QPP model (Figure \ref{fig:overview}).
Here instead of feeding in the interaction tensors between a query and its top-retrieved documents, we simply input the dense vector representations of queries in pairs.
We experiment with two different types of dense vector inputs - one where we used pre-trained RoBERTa vectors \cite{roberta} obtained using the HuggingFace library \cite{web:hugging},
and the other, where we used the sum of the Skipgram \cite{Mikolov13} word embedded vectors (trained on the respective target collections) of constituent terms as the dense representation of a query for input. We name these two ablations as \textbf{SN-BERT} and \textbf{SN-SG}, respectively.    

\para{No Intra-Query Interaction} As another ablation of Deep-QPP, we only use the interaction between the terms of the query pairs themselves. The interaction tensor between a pair of queries is a $2^{nd}$ order tensor, i.e., a $k\times p$ matrix. This is a purely interaction-based method, and in principle, is similar to DRMM \cite{drmm}, with the added layer of 2D convolutions. We denote this baseline as \textbf{DRMM}.   


\subsection{Experiment Settings}
\label{ss:evalmetrics}

\begin{table*}[t]
\centering
\small

\begin{tabular}{l cccccc|cccccc}
\toprule
& \multicolumn{6}{c}{Metric : AP$@100$} & \multicolumn{6}{c}{Metric : nDCG$@20$}
\\

\cmidrule(r){2-7}
\cmidrule(r){8-13}

& \multicolumn{3}{c}{TREC-Robust} & \multicolumn{3}{c}{ClueWeb09B}
& \multicolumn{3}{c}{TREC-Robust} & \multicolumn{3}{c}{ClueWeb09B}
\\

\cmidrule(r){2-4}
\cmidrule(r){5-7}
\cmidrule(r){8-10}
\cmidrule(r){11-13}

Methods  & \multicolumn{1}{c}{Pairwise} & \multicolumn{2}{c}{Pointwise} & 
\multicolumn{1}{c}{Pairwise} & 
\multicolumn{2}{c}{Pointwise} &
\multicolumn{1}{c}{Pairwise} & 
\multicolumn{2}{c}{Pointwise} &
\multicolumn{1}{c}{Pairwise} & 
\multicolumn{2}{c}{Pointwise}
\\

\cmidrule(r){2-4}
\cmidrule(r){5-7}
\cmidrule(r){8-10}
\cmidrule(r){11-13}

& Accuracy & P-$\rho$ & K-$\tau$ 
& Accuracy & P-$\rho$ & K-$\tau$
& Accuracy & P-$\rho$ & K-$\tau$ 
& Accuracy & P-$\rho$ & K-$\tau$
\\

\midrule

Clarity \cite{croft_qpp_sigir02} & 0.6251 & 0.4863 & 0.3140 
& 0.6120 & 0.1911 & 0.0641
& 0.6118 & 0.3529 & 0.2462
& 0.6101 & 0.0923 & 0.0714
\\

NQC \cite{kurland_tois12} & 0.6720 & 0.5269 & 0.4041 
& 0.7030 & 0.2654 & 0.1518
& 0.6689 & 0.4261 & 0.3017
& 0.6916 & 0.3105 & 0.1987
\\

WIG \cite{wig_croft_SIGIR07} & 0.6613 & 0.5440 & 0.4279 
& 0.6829 & 0.2492 & 0.1920
& 0.6629 & 0.3915 & 0.2407
& 0.6710 & 0.2780 & 0.1823
\\

UEF \cite{uef_kurland_sigir10} & 0.6941 & 0.5523 & 0.4154 
& 0.7217 & 0.3162 & 0.1959
& 0.6792 & 0.5029 & 0.3510
& 0.6925 & 0.3320 & 0.1854
\\

\midrule

SN-BERT & 0.6613 & 0.5208 & 0.4169 
& 0.6902 & 0.2317 & 0.1441
& 0.6529 & 0.5023 & 0.3624
& 0.6724 & 0.2241 & 0.1334
\\

SN-SG & 0.6349 & 0.5112 & 0.3987 
& 0.6273 & 0.2110 & 0.1154
& 0.6147 & 0.4736 & 0.3561
& 0.6231 & 0.2049 & 0.1283
\\

DRMM & 0.5871 & 0.4730 & 0.3710 
& 0.6023 & 0.2014 & 0.1141
& 0.5629 & 0.4038 & 0.3119
& 0.6004 & 0.1927 & 0.1201
\\

\wsup \cite{hamed_neuralqpp} & 0.8123 & 0.7215 & 0.5090 
& 0.7727 & 0.5192 & 0.2828
& 0.7973 & 0.5913 & 0.4126
& 0.7614 & 0.3928 & 0.2337
\\

\midrule

Deep-QPP (MDMQ) & 0.7857 & 0.6988 & 0.4981 
& 0.7414 & 0.4636 & 0.2495
& 0.7632 & 0.5649 & 0.3619
& 0.7189 & 0.3509 & 0.2185
\\

Deep-QPP (SDSQ) & 0.7210 & 0.6303 & 0.4018 
& 0.6844 & 0.4208 & 0.2401
& 0.7284 & 0.5112 & 0.3065
& 0.6753 & 0.3124 & 0.2014
\\

Deep-QPP (MDSQ) & 0.8006 & 0.7203 & 0.4989 
& 0.7426 & 0.4840 & 0.2575
& 0.7824 & 0.5601 & 0.3245
& 0.7037 & 0.3518 & 0.2100
\\

Deep-QPP (SDMQ) & \textbf{0.8420} & \textbf{0.7404} & \textbf{0.5434}  
& \textbf{0.8045}  & \textbf{0.5532} & \textbf{0.3130} 
& \textbf{0.8371} & \textbf{0.6315} & \textbf{0.4614} 
& \textbf{0.7903} & \textbf{0.4431} & \textbf{0.2554}
\\

\bottomrule
                       
\end{tabular}
\caption{
\small
A comparison of the QPP effectiveness between \ours, and a set of unsupervised and supervised baselines (shown in the 1$^{st}$ and the 2$^{nd}$ groups, respectively).
The average accuracy and the correlation values (see Section \ref{ss:evalmetrics}) of \ours~over the best performing baseline - \wsup, are statistically significant (t-test with over 97\% confidence).
}
\label{tab:res_table_merge}
\end{table*}

\para{Implementation}
We used the Java API of Lucene 8.8
\cite{lucene}
for indexing and retrieval;
also to implement the existing unsupervised QPP baselines (e.g., for calculating the document and collection statistics). The supervised baseline - \wsup, and our proposed method - \ours, were both implemented in Keras \cite{keras}.
The code for our proposed method is available for research purposes\footnote{https://github.com/suchanadatta/DeepQPP.git}.  

\para{Metrics} 
Recall from Section \ref{ss:training} that the \ours~model can be trained using either the pairwise and the pointwise objectives.
The pointwise test use-case is the standard practice in existing QPP studies, where given a query, a QPP model predicts a score indicative of the retrieval effectiveness. For this use-case, we evaluate the effectiveness of the QPP methods with standard metrics used in the literature: a) \pr~correlation between the AP values of the queries in the test-set and the predicted QPP scores; b) a ranking correlation measure, specifically Kendall's $\tau$ between the ground-truth ordering (increasing AP values) of the test-set queries and the ordering induced by the predicted QPP scores.

In pairwise testing, the network is presented with pairs of queries from the test set, for which it then predicts binary indications of the relative order of queries within the pairs. As a QPP effectiveness measure, we report the average accuracy of these predictions, i.e. whether a predicted relation as given by the $\sigmoid$ output from \ours,
$\hat{y}(Q_a, Q_b;\Theta)$, 
matches the ground-truth that $AP(Q_a) < AP(Q_b)$.
Since $\hat{y}(Q_a, Q_b;\Theta) \in [0,1]$, we \emph{binarize} this value to $\{0,1\}$  with the threshold of $0.5$, thus indicating a prediction of whether $Q_a$ is a more difficult query than $Q_b$ or vice versa.

\para{\ours~hyper-parameters}
For the \ours~method (and also for the semantic analyzer component of the weakly supervised baseline \wsup), we use skip-gram word vectors of dimension $300$ trained on the respective document collections with a window size of $10$ and $25$ negative samples.
Another hyper-parameter in \ours~is the number of intervals (bins) $p$ used to compute the interactions in Equation \ref{eq:interaction-qterm-doc}.
In Table \ref{tab:res_table_merge}, we report results with $p=30$ (as per the settings of the DRMM paper \cite{drmm}), and later investigate the effect of varying this parameter on the effectiveness of \ours~(results in Figure \ref{fig:bin_compare}).

We observed that, after a number of initial experiments, excluding the idf of terms in the interaction tensors always produced worse results than when including them. Therefore, in all our experiments with \ours, we use the idf-weighted interactions (Equation \ref{eq:interaction-qterm-doc}), and do not report the results obtained with Equation \ref{eq:interaction-qterm-doc-noidf} for brevity. Another hyper-parameter that we use in the \ours~model to avoid over-fitting is the dropout probability, which we set to $0.2$ as per the initial trends in our experimental findings.

\subsection{Results} \label{ss:results}
%

%

Table \ref{tab:res_table_merge} presents the QPP results for all the methods investigated. \emph{Firstly}, we observe that the existing supervised approach for QPP, \wsup, outperforms the unsupervised approaches (NQC, WIG and UEF), which conforms to the observations reported in \cite{hamed_neuralqpp}. 
\emph{Secondly}, we observe that the ablation baselines of \ours~involving a purely representation-based approach (SN-BERT and SN-SG), or a purely interaction-based one (DRMM), perform worse than \ours. This is mainly because these baselines lack the additional source of information -- interactions of queries with the top-retrieved set of documents, which \ours~is able to leverage from. This observation also reflects the fact that post-retrieval QPP approaches, with the additional information from top-documents, typically outperform pre-retrieval ones \cite{kurland_tois12}.

\emph{Third and most importantly}, we observe that \ours~outperforms \wsup, which 
confirms the hypothesis that explicitly learning the relative specificity of query pairs with an end-to-end (strongly) supervised model is better able to generalize than a weakly supervised approach which learns an optimal combination of statistical predictors. 

Another observation is that the SDMQ version of the reshaping function $\pi: \mathbb{R}^3\mapsto \mathbb{R}^2$ (see Section \ref{ss:reshape} and Figure \ref{fig:overview}) turns out to be the most effective, as we might expect. This also conforms to the way in which unsupervised QPP approaches generally work, i.e., by first making use of the information from each top-retrieved document (e.g. its score in NQC and WIG) and then computing an aggregate function over them (e.g. their variance in NQC, and relative gains in WIG).  

\begin{figure}[!ht]
\centering
\includegraphics[width=\columnwidth]{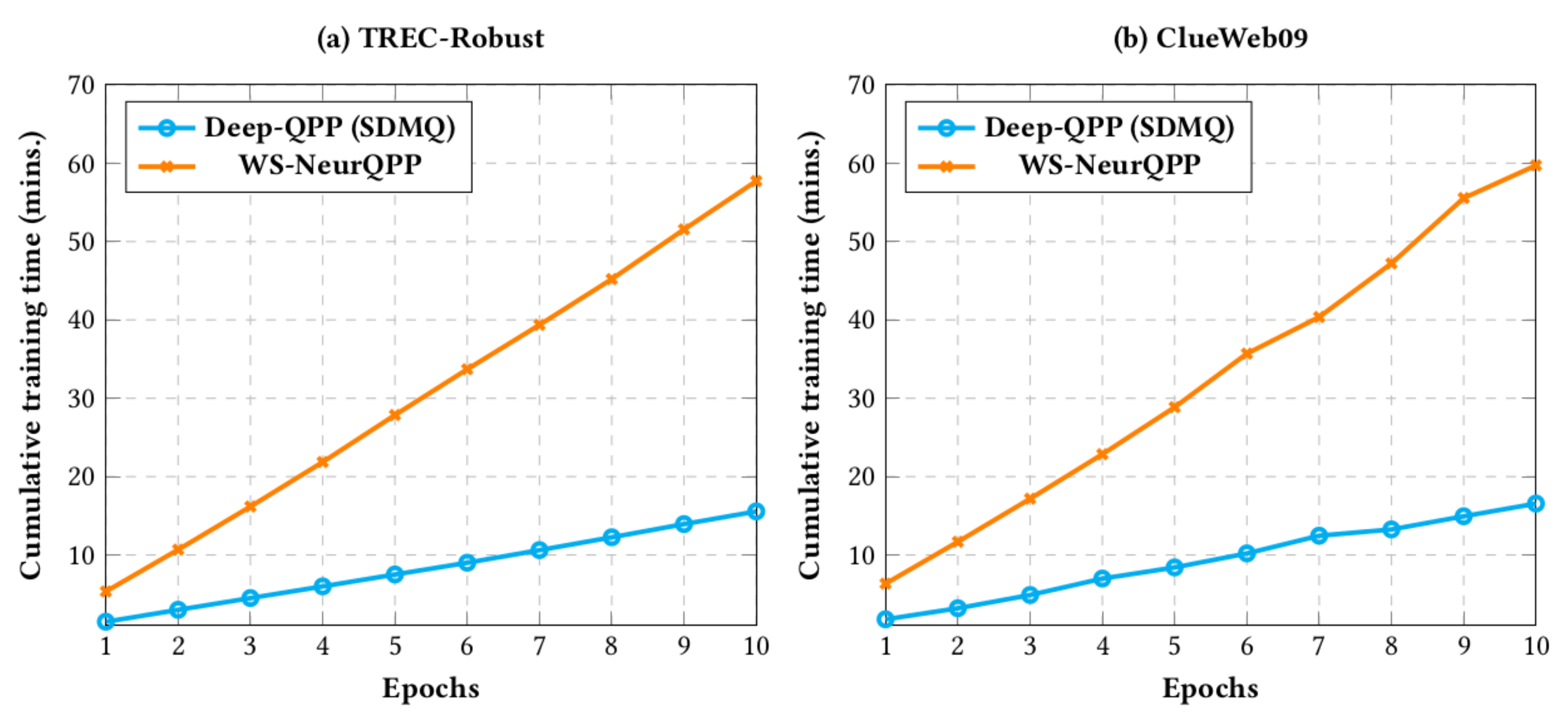}
\caption{
\small
\ours, in addition to being more effective than \wsup, also outperforms \wsup~in terms of training time because of a much smaller number of parameters (1.9M vs. 4.7M).}
\label{fig:epoch_vs_time}
\end{figure}

To further compare \ours~ to \wsup, we report the training-time efficiency of both approaches in Figure \ref{fig:epoch_vs_time}.
Due to a much larger number of trainable parameters and larger input dimensionality (dense word vectors instead of interactions between the dense vectors), \wsup~turns out to be taking a much larger time to execute than \ours.  
The total number of trainable parameters of~\wsup~is $4.7$M which is about $2.5$X the number of parameters in \ours~($1.9$M).

\begin{figure}[!ht]
\centering
\includegraphics[width=\columnwidth]{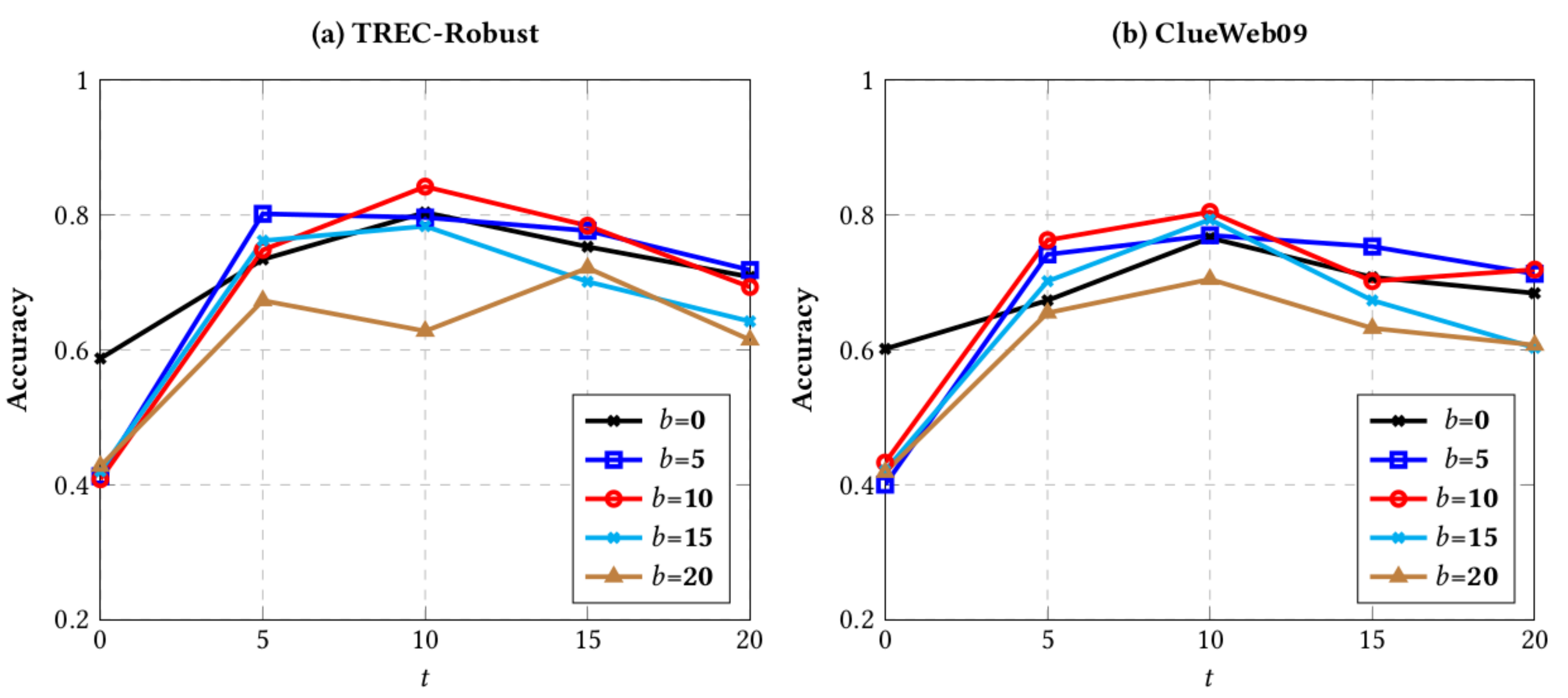}
\caption{
\small
Sensitivity of \ours~on the number of top ($t$) and bottom ($b$) documents to include for interaction computation (see Section \ref{ss:qd-interaction}) on QPP effectiveness. The limiting case of $(t, b)=(0, 0)$ corresponds to the situation when we simply use the interaction between query terms themselves (i.e. the DRMM baseline).}
\label{fig:t-b-accu}
\end{figure}

\para{Hyper-parameter Sensitivity of \ours}
Figure \ref{fig:t-b-accu} shows that using the top-$10$ and the bottom-$10$ documents for the interaction computation (Section \ref{ss:qd-interaction}) yields the best results, which shows that neither a too small nor too large a number of documents should be used as inputs for learning the QPP comparison function.

\begin{figure}[!ht]
\centering
\includegraphics[width=\columnwidth]{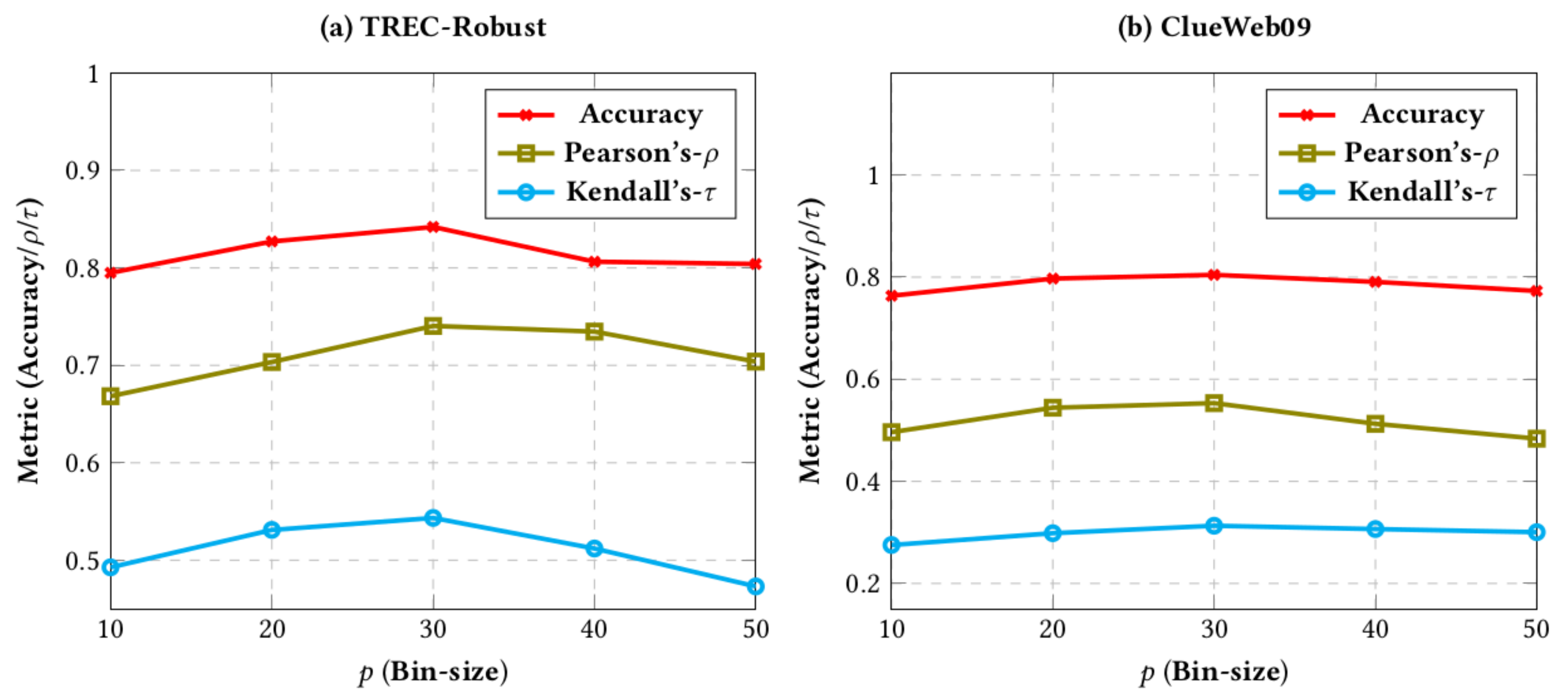}
\caption{
\small
Sensitivity of \ours~w.r.t. the bin-size, $p$.
}
\label{fig:bin_compare}
\end{figure}

Figure \ref{fig:bin_compare} shows the effects of different bin-sizes, $p$ (of Equation \ref{eq:interaction-qterm-doc}), used to compute the interactions between queries and the documents retrieved at top and bottom ranks. A value of $30$ turned out to be optimal, which is similar to the reported optimal value of the bin-size for interaction computation in the LTR task \cite{drmm}.


\section{Related Work} \label{sec:review}
We have already discussed a number of existing QPP methods as a part of the description of the baselines in Section \ref{ss:baselines}.
We now outline additional QPP work, and also cover some recent work on applications of end-to-end learning in IR.
%
Kurland et. al. \cite{qpp-cluster} showed that the QPP task is equivalent to ranking clusters of similar documents by their relevance with respect to a query. Zendel et. al. \cite{query_variants_kurland} made use of alternative expressions of information needs, such as variants of a given query, to improve QPP effectiveness.
The study \cite{fernando_correlation_sigir07} reported that a spatial analysis of vector representations of top-retrieved documents provide useful cues for improving QPP effectiveness -- a hypothesis that our data-driven model also includes, through the convolutions over the interaction matrices.
Other standard deviation-based approaches, somewhat similar to NQC, have also been reported to work well for the QPP task \cite{cummins_qppSD_sigir11,cummins}.
Apart from the weakly supervised neural approach of \wsup~\cite{hamed_neuralqpp}, a QPP unsupervised approach that uses cluster hypothesis of word vectors in an embedded space was proposed in \cite{qpp_DG_IPM}.

Recent studies have reported a close association between the findings of learning to rank (LTR) and QPP studies. It was reported that the set of features that are useful for LTR also proves beneficial for QPP \cite{qpp_letor,Djean2020ForwardAB}. Moreover, the mechanism of two levels of interaction (both between queries and documents, and across queries) has also been reported to be useful for LTR \cite{mitra-www}.   

In addition to DRMM \cite{drmm}, other work proposing end-to-end LTR approaches include \cite{knrm_SIGIR17,neural_rank_multi_doc_field_www-18}. The ColBERT model was recently proposed in \cite{colbert_sigir20}, which is a fine-tuned BERT model \cite{devlin-etal-2019-bert} using pairwise ranking loss. As a precursor to end-to-end supervised approaches, unsupervised approaches have addressed term semantics by using dense word vectors, including \cite{GangulyRMJ15,kdrelm_DG,dg-cikm-we} which used skip-gram vectors and the work of \cite{birch} which employed BERT.

\section{Conclusions and Future work}
\label{sec:concl}
In this paper, we have proposed \ours, a data-driven end-to-end neural framework for the task of query performance prediction in ad-hoc retrieval.
Rather than relying on statistical term weighting heuristics or employing a weakly-supervised model on those heuristics, our method directly learns from the data, where the input consists of a set of queries, along with their top-retrieved sets of documents. 
The ground-truth for training is comprised of the true query performance indicators (e.g., measured with AP).
\balance Our experiments, conducted on standard news and Web collections, demonstrated that a data-driven approach trained on query pairs with known QPP indications (e.g., AP values) is able to effectively generalize this comparison function for unseen query pairs. The improvement percentages obtained for Web queries are in fact higher which suggest that, in future we could potentially use pseudo-relevance information in the context of query logs, such as clicks and dwell times, to train QPP models at a large scale.


\subsubsection*{\textbf{Acknowledgement}}
The first and the third authors were partially supported by the Science Foundation Ireland (SFI) grant number SFI/12/RC/2289\_P2. 

\bibliographystyle{ACM-Reference-Format}
\bibliography{final_refs}

\end{document}